\documentclass[prb,twocolumn,showpacs,amsmath,amssymb]{revtex4}
\usepackage{graphicx}
\usepackage{graphicx}% Include figure files

\begin{document}

\title{Varying Eu$^{2+}$ magnetic order in EuFe$_2$As$_2$ by chemical pressure}

\author{S. Zapf}
%\affiliation{1.~Physikalisches Institut, Universit\"at Stuttgart, Pfaffenwaldring 57, 70550 Stuttgart, Germany}
\author{D. Wu}
\affiliation{1.~Physikalisches Institut, Universit\"at Stuttgart, Pfaffenwaldring 57, 70550 Stuttgart, Germany}
\author{L. Bogani}
\affiliation{1.~Physikalisches Institut, Universit\"at Stuttgart, Pfaffenwaldring 57, 70550 Stuttgart, Germany}
\author{M. Dressel}
\affiliation{1.~Physikalisches Institut, Universit\"at Stuttgart, Pfaffenwaldring 57, 70550 Stuttgart, Germany}
\author{H. S. Jeevan}
\affiliation{I.~Physikalisches Institut, Universit\"at G\"ottingen, Friedrich-Hund-Platz 1, 37077 G\"ottingen, Germany}
\author{P. Gegenwart}
\affiliation{I.~Physikalisches Institut, Universit\"at G\"ottingen, Friedrich-Hund-Platz 1, 37077 G\"ottingen, Germany}

%\email{dressel@pi1.physik.uni-stuttgart.de}
%\homepage{http://www.pi1.physik.uni-stuttgart.de}
\date{\today}

\begin{abstract}
Among iron 122 pnictide superconductors, the EuFe$_2$As$_2$
series draws particular interest because, in addition to
superconductivity or the long-range spin-density-wave order in the
Fe subsystem, the localized Eu$^{2+}$ magnetic moments order at
low temperatures. Here we present a novel scheme of how the spins
align in the Eu compounds when pressure varies the coupling; we
explain magnetization measurements on
EuFe$_2$(As$_{1-x}$P$_x$)$_2$ single crystals as well as other
observations of the Eu$^{2+}$ ordering previously reported in
literature. The magnetic moments of the Eu$^{2+}$ ions are
slightly canted even in the parent compound EuFe$_2$As$_2$,
yielding a ferromagnetic contribution along the $c$-direction that
becomes stronger with pressure. Reducing the interlayer distance
even further, the antiferromagnetic coupling of the $ab$ planes
finally turns ferromagnetic.
\end{abstract}

%%%%%%%%%%%%%%%%%%%%%%%%%%%%%%%%%%%%%%%%%%%%%%%%%%%%%%%%%%%%%%%%%%%%%%%%%%%%%%%%%%%%%%%%%%%%%%%%%
%%%%%%%%%%%%%%%%%%%%%%%%%%%%%%%%%%%%%pacs
\maketitle

With the discovery of superconductivity  in iron pnictides, a new
approach was taken towards understanding the mechanism of
high-temperature superconductivity. Magnetism seems to be the
crucial factor determining the physical properties, as
superconductivity emerges with the suppression of the
spin-density-wave (SDW) ordering in the FeAs
layers.\cite{Johnston10} Although the highest critical
temperatures $T_c$ are reached in the quaternary 1111 compounds,
%(Gd$_{0.8}$Th$_{0.2}$FeAsO with $T_c = 56$~K, Ref.~\onlinecite{Wang08})
the ternary 122 compounds
%(with maximum $T_c = 38$~K in Ba$_{0.6}$K$_{0.4}$Fe$_2$As$_2$, Ref.~\onlinecite{Rotter08})
quickly advanced as the model systems of iron pnictides due to the
availability of good single crystals. Besides the parent compounds
BaFe$_2$As$_2$ and SrFe$_2$As$_2$, EuFe$_2$As$_2$ is an
outstanding member of the 122 series with a maximum $T_c=32$~K in
Eu$_{0.5}$K$_{0.5}$Fe$_2$As$_2$,\cite{Jeevan08} because in
addition to the antiferromagnetic (AFM) order of the itinerant
electrons in the FeAs layers at $T_{\rm SDW}=189$~K, magnetic
order of the localized Eu$^{2+}$ ions is observed at low
temperatures. Eu$^{2+}$ possesses a large magnetic moment ($J =
7/2$) that leads to a so called ``A-type'' AFM ordering below $T_N
= 19$~K,\cite{Jiang09,Herrero09,Xiao09} meaning that the Eu$^{2+}$
moments align ferromagnetically along the $a$-axis and
antiferromagnetically along the $c$-axis. The coexistence and
interplay of the second magnetic ordering with the SDW in the FeAs
layers is drawing increasing attention.

Charge-carrier doping on the Eu- or Fe-sites,
\cite{Ren09,Jiang09b,Jeevan08} isovalent P substitution
on the As site\cite{Ren09b}  and physical pressure
\cite{Miclea09,Terashima09} change the Eu$^{2+}$ ordering as well
as the electronic properties of the system, eventually leading to
superconductivity. One interesting signature of these
superconductors is a resistivity re-entrance below $T_c$ in the
range of the Eu$^{2+}$ ordering temperature, observed for example
in Eu(Fe$_{0.89}$Co$_{0.11}$)$_2$As$_2$,\cite{Jiang09b}
EuFe$_2$(As$_{0.7}$P$_{0.3}$)$_2$,\cite{Ren09b} and EuFe$_2$As$_2$
under pressure.\cite{Miclea09,Terashima09} The re-entrance is
suppressed by a magnetic field applied in the $ab$-plane, but not affected by a field parallel to the
$c$-axis.\cite{Jiang09b} This implies that the Eu$^{2+}$ 
ordering has noticeable impact on the superconductivity in the
FeAs layers and that the Eu 122 compounds reveal an extraordinary
possibility to study the interplay between magnetism and
superconductivity.

It is therefore quite surprising that, at present, no clear picture
exists how the ordering of the Eu spins changes with doping or pressure. Furthermore, it is still under
debate which kind of Eu$^{2+}$ magnetic ordering coexists with
superconductivity and different phase
diagrams have been proposed even for the same compound.\cite{Jeevan11,Nowik11} In the
following we concentrate on EuFe$_2$(As$_{1-x}$P$_x$)$_2$ and
discuss the influence of chemical pressure on the alignment of the
Eu$^{2+}$ magnetic moments. We suggest that the Eu spins are
slightly canted along the $c$-axis, causing an appreciable
ferromagnetic (FM) contribution that increases with pressure. Our
scheme is consistent with published data and explains the apparent
discrepancies in phase diagrams of Eu magnetic
ordering.

\section{Experimental results}
EuFe$_2$(As$_{1-x}$P$_{x}$)$_2$ single crystals used in this
study were synthesized by a Bridgman method and characterized as
previously described.\cite{Jeevan11} Optical investigations on crystals with $x = 0$ and $x = 0.18$ have been already reported.\cite{Wu09,Wu11} Here we describe the magnetic behavior of EuFe$_2$As$_2$ and
EuFe$_2$\-(As$_{0.88}$P$_{0.12}$)$_2$, measured with a
Quantum Design MPMS-XL superconducting quantum interference device
(SQUID). We provide a complete characterization as a function of temperature ($T$) and magnetic field ($H$) for the main crystallographic directions. EuFe$_2$As$_2$ has orthorhombic
symmetry with $a$ and $b$ axes virtually identical. Even though twinning of the crystals did not allow us a characterization in the $ab$-plane, neutron scattering data indicate that the Eu$^{2+}$ moments align along the $a$-direction. 
\begin{figure}
 \centering
  \includegraphics[width=0.6\columnwidth]{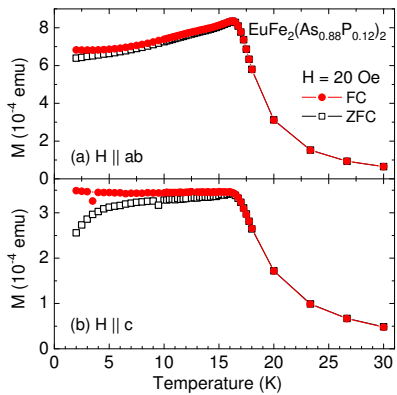}
  \caption{Zero-field cooled (ZFC) and field-cooled (FC) magnetization curves (solid red circles and open black squares, respectively) for EuFe$_2$(As$_{0.88}$P$_{0.12}$)$_2$, measured in $H=20$~Oe parallel (a) and perpendicular (b) to the $ab$-plane.}
  \label{G_T}
\end{figure}
In Fig.~\ref{G_T} the temperature-dependent magnetization of
EuFe$_2$(As$_{0.88}$P$_{0.12}$)$_2$ is plotted for $H=20$~Oe, measured parallel and perpendicular to the $ab$-plane. While at elevated temperatures ($T<T_{\rm SDW}$) the
paramagnetic regime can be nicely described by the Curie-Weiss
law, similar to the parent compound,\cite{Wu09} the cusp at
$T_N=16$~K evidences AFM order of the  Eu$^{2+}$ moments. At low
temperatures we find a distinct difference between the
field-cooled (FC) and zero-field cooled (ZFC) behavior (especially for
$H\parallel c$) that seems to be more pronounced as P substitution
increases.\cite{Jeevan11}

\begin{figure}
 \centering
  \includegraphics[width=0.6\columnwidth]{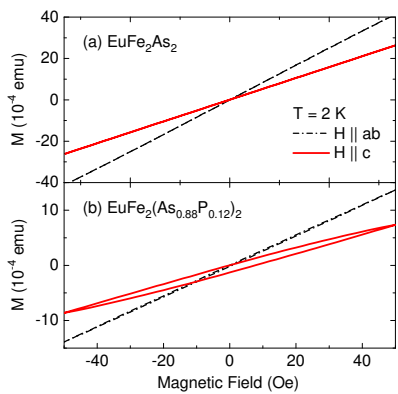}
  \caption{Isothermal magnetization {\em versus} external magnetic field applied parallel (black dashed lines) and perpendicular (red solid line) to the $ab$-plane for $T=2$~K.
(a) While for EuFe$_2$As$_2$ no hysteresis  can be observed, (b)
in the case of EuFe$_2$(As$_{0.88}$P$_{0.12}$)$_2$ the hysteresis
is obvious for $H\parallel c$ but can also be identified for
$H\parallel ab$.}
  \label{G_Hyst}
\end{figure}
\begin{figure}
    \centering
        \includegraphics[width=0.9\columnwidth]{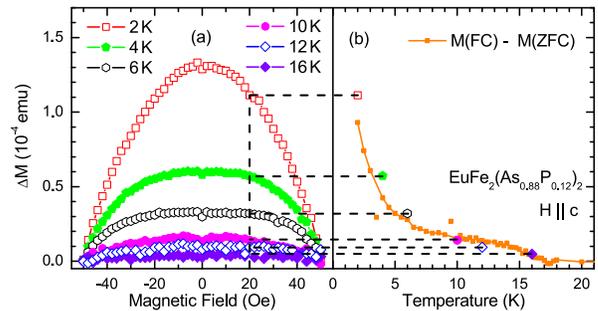}
    \caption{(a) Height of the hysteresis for EuFe$_2$(As$_{0.88}$P$_{0.12}$)$_2$, measured with $H\parallel c$ for different fields and temperatures. (b) Temperature dependence of the difference between the ZFC and FC curves shown in Fig \ref{G_T}(b). This difference can be compared to the hysteresis height at $H=$ 20~Oe shown in (a). The broken lines are a guide to the eye to stress the correlation between the two trends.}
    \label{G_Hyst_T}
\end{figure}

To clarify this point, we measured the field dependence of
the magnetization parallel and perpendicular to the $ab$-plane. The complete magnetization {\em versus} field curve has already been discussed\cite{Jiang09, Nowik11} and here we concentrate on the low-field features.
Fig.~\ref{G_Hyst} displays the results obtained at $T=2$~K for
EuFe$_2$As$_2$ and EuFe$_2$(As$_{0.88}$P$_{0.12}$)$_2$. While we
observe no hysteresis for the parent compound,
EuFe$_2$(As$_{0.88}$P$_{0.12}$)$_2$ exhibits clear evidence of FM
behavior when the magnetic field is applied parallel to the
$c$-axis. For $H\parallel ab$ only a much smaller hysteresis is observed. 
The hysteresis becomes narrower with increasing
temperature and completely  vanishes at 16~K
(Fig.~\ref{G_Hyst_T}), which is the magnetic transition
temperature common to both crystal directions. To follow the
opening of the hysteresis, we plot in Fig.~\ref{G_Hyst_T} the difference $\Delta M$ between the magnetization curves acquired by sweeping the field down and up ($i.e.$ the hysteresis height). The curves show a decrease of the hysteresis height on raising the temperature and it is instructive to compare this trend with the difference between the ZFC and FC curves at the same field. As shown in Fig.~\ref{G_Hyst_T}(b) the two trends coincide, demonstrating the development of magnetic ordering.

\section{Model}
\begin{figure}[t]
 \centering
  \includegraphics[width=0.85\columnwidth]{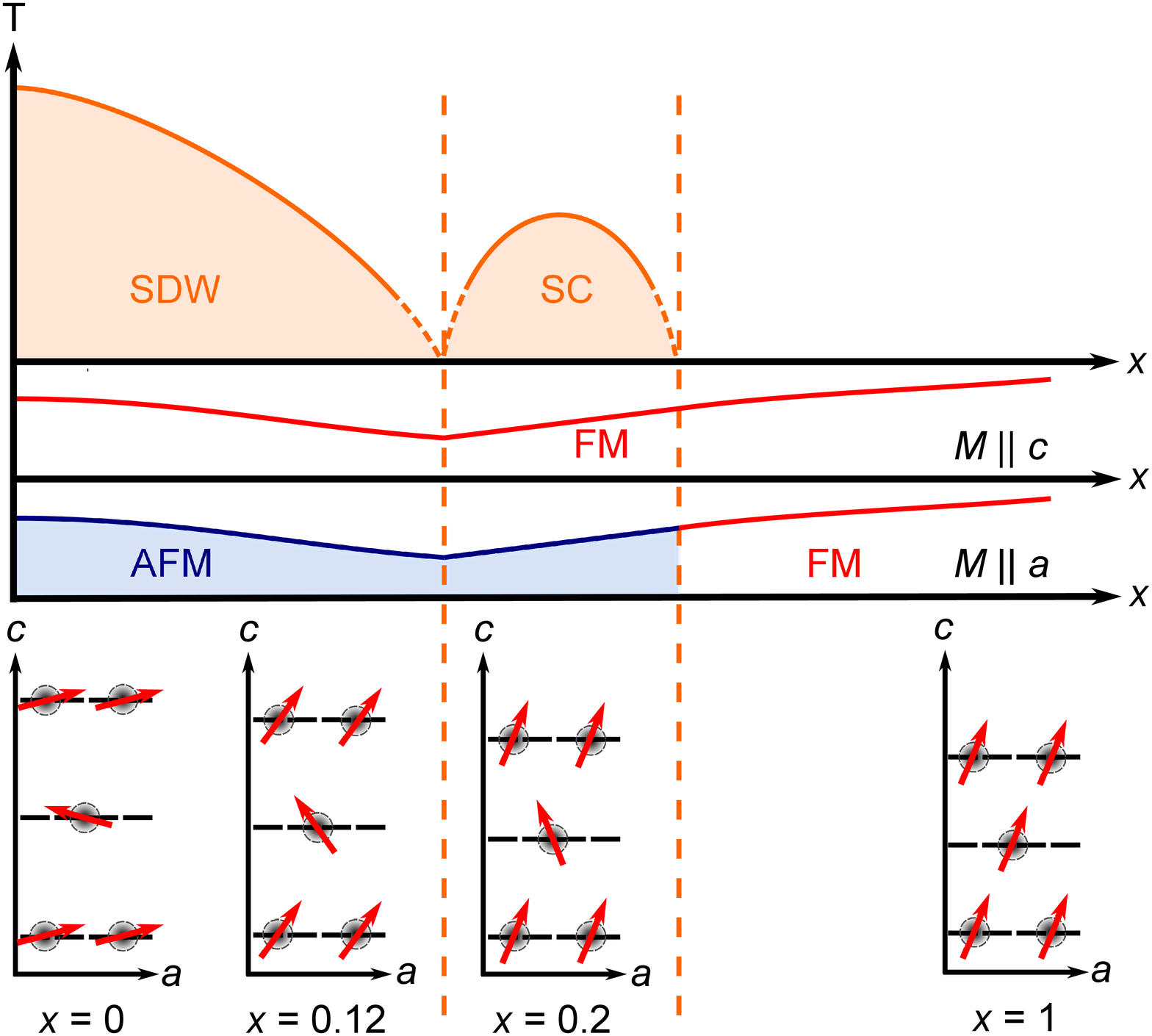}
   \caption{Schematic phase diagram of
   EuFe$_2$(As$_{1-x}$P$_x$)$_2$  indicating the SDW
   and SC phase. In the preceding panels, we
    distinguish between the Eu net magnetization along the $a$-axis
    ($M\parallel a$) and along the $c$-axis ($M\parallel c$) resulting
    from the present data and from Ref.~\onlinecite{Jeevan11,Nowik11}.
    The scheme of the Eu$^{2+}$ spin alignment is also shown.
    In the parent compound, A-type AFM is found with the spins being slightly canted towards the $c$-axis with an
    angle\cite{Nowik11} of 13$^\circ$ to the $ab$-plane. Upon the application of
    pressure the canting angle increases until it saturates\cite{Nowik11} at 68$^\circ$ with respect to the $a$-axis ($x = 0.2$). At even higher pressure, the AFM interlayer coupling turns probably into a FM one.}
    \label{PhasenDiagrammNeu}
\end{figure}

Based on these magnetization measurements and in agreement with other
studies, we propose that the magnetic
moments of the Eu$^{2+}$ ions change their alignment under
pressure as displayed in Fig.~\ref{PhasenDiagrammNeu}. In
the parent compound, the spins orient along the $a$-direction, but basically reverse the direction in adjacent
$ab$-planes.\cite{Jiang09} More precisely, one observes canted
AFM causing a small FM component of the moments in $c$-direction that
is usually neglected in the discussion. M\"ossbauer
studies reveal a steep increase of this canting angle with
increasing P substitution
until the Eu spins are aligned  almost perpendicular to the
$ab$-plane.\cite{Nowik11} We want to emphasize that this
has a huge impact on magnetic susceptibility and has to
be considered in the interpretation.

Due to the non-negligible tilt angle out of the $ab$-plane, as
displayed in Fig.~\ref{PhasenDiagrammNeu}, one always observes a FM
signal along the $c$-axis,
including a hysteretic behavior. This is exactly what we measured
in our EuFe$_2$(As$_{0.88}$P$_{0.12}$)$_2$ single crystal. The small hysteresis visible for $H\parallel
ab$ is likely caused by our limited precision in crystal alignment and we will not discuss it further. The
fact that in the parent compound the hysteresis cannot be resolved
might be due to the extremely small spin canting. Since
chemical and physical pressures have a similar effect
on the 122 compounds, we propose to examine  EuFe$_2$As$_2$ single
crystals by a high-pressure SQUID susceptometer to
confirm and  complete the phase diagram. These measurements will
reveal  whether at higher pressure the AFM interlayer coupling finally turns into a FM one, as claimed for different Eu 122
compounds.\cite{Ren09,Jeevan11,Nowik11}  In Fig.~\ref{PhasenDiagrammNeu} we also sketch the arrangement of the Eu$^{2+}$ moments in the high-pressure limit ($x = 1$).

\section{Discussion}
The hysteresis loop observed in our
EuFe$_2$\-(As$_{0.88}$P$_{0.12}$)$_2$ single crystal measurements
constitute the key to understand the various phase diagrams of
Eu 122 pnictides proposed in literature. Since contributions from
the $c$-direction are unavoidable in polycrystalline samples, a
hysteresis is always seen as soon as the canting of the spins is
pronounced enough. In the following we compare the phase
diagram of Jeevan {\it et al.} based on single crystal
measurements\cite{Jeevan11} with that of Nowik {\it et al.} who
performed measurements on polycrystalline samples.\cite{Nowik11}
The former observe AFM Eu$^{2+}$ ordering in the superconducting
phase, whereas the latter group claims the coexistence of FM and
superconductivity. The statements do not contradict each other
since Jeevan {\it et al.} refer to the ordering of the Eu$^{2+}$
moments in the $ab$-plane. If they measure along the $c$-axis,
they will also yield a phase diagram with the Eu ordering along
the $c$-axis; for this direction superconductivity coexists with
FM. We sketch such a phase diagram in Fig.~\ref{PhasenDiagrammNeu},
where we distinguish between the Eu net magnetization along the
$a$- and $c$-axis. This causes implications that are in particular
interesting for theoretical considerations about the interplay of
magnetism and superconductivity, as superconductivity in pnictides
is often considered as quasi-two-dimensional.

On the basis of in-plane magnetization measurements of
EuFe$_2$\-(As$_{1-x}$P$_x$)$_2$ single crystals, Jeevan {\it et
al.}\cite{Jeevan11} proposed a phase diagram that includes FM
Eu$^{2+}$ ordering for high P content. Assuming that the
sample was perfectly oriented with $H\parallel ab$, this suggests the
spin alignment displayed in Fig.~\ref{PhasenDiagrammNeu} for $x = 1$. This is supported by the fact that the hysteresis continues to open  with increasing P
substitution,\cite{Jeevan11} while the canting angle
already saturates at $x = 0.2$.\cite{Nowik11}

There is still some discussion about the actual coupling mechanism
between the Eu layers. Ruderman-Kittel-Kasuya-Yosida (RKKY) interaction might play an important
role (although it is not clear what happens in the superconducting
state), but other mediation via the Fe layers is also possible,
since direct exchange and dipolar interaction can be ruled out.
When going from EuFe$_2$As$_2$ to EuAs$_2$P$_2$ the $a$ and $c$
axes decrease by 2.3\% and 7.3\%, respectively.\cite{Feng10} The reduction of
the unit-cell  volume and the change of the Fermi surface with
pressure\cite{Wu11,Thirupathaiah11} could turn an interlayer RKKY coupling  from AFM to FM.\cite{Ren09}

Field-dependent magnetization measurements ranging up to several
Tesla can be also explained by the Eu$^{2+}$ spin alignment
displayed in Fig.~\ref{PhasenDiagrammNeu} and indicate FM interlayer coupling at higher pressure. In EuFe$_2$As$_2$ the
magnetic behaviors for $H\parallel ab$ and  $H\parallel c$
differ significantly.\cite{Jiang09} While in both directions
saturation is achieved around 1~T, a step-like magnetization curve
appears only in the case of $H\parallel ab$. At low field
strength, the Eu$^{2+}$ spins start to realign, but keep their AFM
orientation in adjacent planes; above a kind of spin-flip field
the orientation of the spins parallel to the external field wins
over the AFM interlayer coupling. In the case of $H\parallel c$,
however, we propose that such magnetization behavior does not
appear since there is no AFM ordering in the $c$-direction that
has to be broken (Fig.~\ref{PhasenDiagrammNeu}).
According to our model, the step-like behavior of the
magnetization with $H\parallel ab$ should become weaker as the canting angle of the
Eu$^{2+}$ spins out of the $ab$-plane grows with pressure. For
larger canting angles, it gets more and more energetically
favorable to orient the spins directly parallel to the external
field instead of keeping the AFM ordering. In the case of the FM
ordered phase (Fig.~\ref{PhasenDiagrammNeu}, $x = 1$), the
step-like behavior vanishes. Actually, this
magnetization dependence was observed for
EuFe$_{2-x}$Ni$_x$As$_2$, with a weak step in the
case of $x = 0.03$, which disappears for higher
dopings.\cite{Ren09}

Pressure-dependent susceptibility measurements would help to identify the correlations between the Eu$^{2+}$ ordering and other phase transitions.
It is quite surprising that the tilting angle of the Eu$^{2+}$ moments stays the same, when the SDW is completely suppressed (Fig.~\ref{PhasenDiagrammNeu}).\cite{Nowik11} This infers that the itinerant AFM phase of the FeAs layers, with the magnetic moments being aligned within the $ab$-plane,\cite{Johnston10} does influence the Eu$^{2+}$ ordering. It is unlikely that the minor structural phase transition between the orthorhombic and tetragonal phase, taking place at approximately the same P content, has some influence on the canting angle.

Understanding the alignment of the Eu$^{2+}$ magnetic moments will also
reveal more information on the interplay between magnetism
and superconductivity. One interesting feature appearing in the Eu
122 pnictides is a resistivity re-entrance in the superconducting
phase near the Eu$^{2+}$ magnetic ordering temperature.\cite{Jiang09b,Ren09b,Jeevan11}
In Eu(Fe$_{0.89}$Co$_{0.11}$)$_2$As$_2$ the Eu$^{2+}$ are ordered AFM in the $a$-direction. The superconducting
re-entrance can be suppressed with an
external magnetic field in the $ab$-plane, while a magnetic field
$H\parallel c$ has no influence on the re-entrance.\cite{Jiang09b}
We do not understand yet, why the AFM Eu$^{2+}$ ordering
destroys the superconductivity, while it can coexist with the field-induced FM.
Nevertheless our model allows us to explain why
a magnetic field applied along the $c$-direction has no appreciable influence: it
only weakens the spin components in the $ab$-plane without destroying
the AFM ordering itself. It should also be noted that for several
Eu 122 compounds, susceptibility measurements show additional
features below the pronounced AFM or FM Eu ordering, which could
have the same origin as the resistivity re-entrance.\cite{Jiang09b, Nowik11, Ren09} Here we want
to point out that Ahmed {\it et al.}\cite{Ahmed10} observe a significant
enhancement of the spin moment on the Fe $3d$ electrons in
EuFe$_2$(As$_{0.73}$P$_{0.27}$)$_2$ at $T=18$~K. Together with
the ordered Eu$^{2+}$ moments, the internal magnetic field can then exceed the superconducting upper critical field leading to the re-entrance of resistivity.

\section{Concluding Remarks}
Analyzing measurements of the magnetic properties of various Eu
122 compounds we propose that the alignment of the localized
Eu$^{2+}$  magnetic moments in EuFe$_2$\-(As$_{1-x}$P$_x$)$_2$
changes with pressure. In the parent compound
EuFe$_2$As$_2$, the spins are not simply antiferromagnetically
aligned in adjacent $ab$-planes but canted by a few degrees,
causing a small ferromagnetic component in $c$-direction. When the
interlayer coupling changes with pressure, the canting increases until the SDW in the FeAs
subsystem is suppressed and superconductivity sets in. There is a
coexistence of superconductivity with antiferromagnetism along the
$a$-direction and ferromagnetism along the $c$-axis. At even
higher pressure, the interlayer coupling switches to a ferromagnetic
one.
\\
\begin{acknowledgments}
We thank: N. Bari\v{s}i\'{c} and R. Beyer  for helpful discussions; the
Alexander von Humboldt Foundation, DFG (SPP1458) and the
Lan\-des\-gra\-du\-ier\-ten\-f\"or\-derungs\-ge\-setz
Baden-W\"urttemberg for financial support. 
\end{acknowledgments}

\end{document}